\documentstyle[12pt,twoside,fleqn,espcrc1,epsf]{article}

\newcommand{\be}{\begin{eqnarray}}
\newcommand{\ee}{\end{eqnarray}}

\newcommand{\AmS}{{\protect\the\textfont2
  A\kern-.1667em\lower.5ex\hbox{M}\kern-.125emS}}

\title{Color superconductivity: Continuity of quark and hadron matter, 
the role of the strange quark mass, and perturbative results}

\author{Thomas Sch\"afer\address{School of Natural Sciences\\
Institute for Advanced Study\\
Princeton, NJ 08540}}

\begin{document}
% typeset front matter
\maketitle

\begin{abstract}
  We summarize some recent results on the structure of QCD at very 
high baryon density.
\end{abstract}

 \section{Introduction}

 It was pointed out almost 20 years ago that asymptotic freedom
and the presence of a Fermi surface imply that QCD at very large
density is a color superconductor \cite{BL_84}. This result
was largely forgotten until it was realized that, using 
interactions that reproduce the strength of chiral symmetry
at zero density, one expects the superconducting gap to 
be quite large, on the order of $\Delta\simeq 100$ MeV at 
densities $\rho\sim 5\rho_0$ \cite{RSSV_98,ARW_98}. A lot
of work on the superconducting phase of QCD has appeared 
over the past year. In this contribution we would like to 
summarize a number of interesting results. 

 \section{Continuity of quark and hadron matter}

 For two flavors the order parameter has the structure 
$\langle\epsilon^{ab3}\psi^{a\,T}C\gamma_5\tau_2\psi^b\rangle$.
This order parameter leaves the chiral $SU(2)_L\times SU(2)_R$ 
symmetry unbroken, but breaks $SU(3)$ color down to $SU(2)$. 
In the case of three flavors an interesting new possibility 
arises. The order parameter \cite{ARW_98b}
\be
\label{cfl}
\langle\psi^{a\,T}_iC\gamma_5\psi^b_j\rangle
 = \Delta_1 \delta^a_i\delta^b_j
  +\Delta_2 \delta^a_j\delta^b_i
\ee
locks the local color orientation $a,b$ to the flavor orientation
$i,j$. In the color-flavor-locked state color $SU(3)$ is completely
broken and all gluons acquire a mass. In addition to that, the 
chiral $SU(3)_L\times SU(3)_R$ is broken to the diagonal $SU(3)_{L+R+C}$.
This provides a very unusual mechanism for chiral symmetry breaking. 
The chiral structure of the order parameter is $\psi_L\psi_L-\psi_R
\psi_R$, so there is no direct coupling between left and right 
handed fields. Chiral symmetry is broken because color locks the 
residual flavor symmetry of the left handed quarks to the corresponding
symmetry of the right handed quarks. 

 Not only is chiral symmetry broken, but also the spectrum of 
excitations in the color-flavor-locked (CFL) phase looks remarkably 
like the spectrum of QCD at low density \cite{SW_98b}. The excitations 
can be classified according to their quantum numbers under the 
unbroken $SU(3)$, and by their electric charge. The modified
charge operator that generates a true symmetry of the CFL phase
is given by a linear combination of the original charge operator
$Q_{em}$ and the color hypercharge operator $Q={\rm diag}(-2/3,-2/3,1/3)$.
Also, baryon number is only broken modulo 2/3, which means that 
one can still distinguish baryons from mesons. We find that the 
CFL phase contains an octet of Goldstone bosons associated with
chiral symmetry breaking, an octet of vector mesons, an octet 
and a singlet of baryons, and a singlet Goldstone boson related 
to superfluidity. All of these states have integer charges.  

 With the exception of the $U(1)$ Goldstone boson, these states
exactly match the quantum numbers of the lowest lying multiplets
in QCD at low density. In addition to that, the presence of the 
$U(1)$ Goldstone boson can also be understood. The $U(1)$ order
parameter is $\langle (uds)(uds)\rangle$. This order parameter
has the quantum numbers of a $J^\pi=0^+$ $\Lambda\Lambda$ pair 
condensate. In $N_f=3$ QCD, this is the most symmetric two nucleon 
channel, and a very likely channel for superfluidity to occur in 
nuclear matter at low to moderate density. We conclude that in QCD 
with three degenerate light flavors, there is no fundamental difference 
between the high and low density phases. This implies that the
low density nuclear phase and the high density quark phase
might be continuously connected, without an intervening phase
transition. 

 The order parameter (\ref{cfl}) breaks the chiral $SU(3)$ 
symmetry but leaves a discrete $Z_2$ symmetry unbroken. This 
symmetry, however, is explicitly broken by instantons. The
quark condensate $\langle\bar\psi\psi\rangle$ also violates 
$Z_2$ symmetry. In weak coupling, the quark condensate is
therefore generated by instantons. This is easy to see from
the structure of the instanton induced interaction between
quarks. In the case $N_f=3$, the instanton vertex is of the
form $(\bar\psi_L\psi_R)^3$. In the CFL phase, we can 
absorb four of the external legs into the $\langle\bar\psi_L
\bar\psi_L\rangle$ and $\langle\psi_R\psi_R\rangle$ 
condensates. The remaining $\bar\psi_L\psi_R$ vertex 
directly generates a quark condensate. The magnitude of 
the chiral condensate in the CFL phase was calculated in
\cite{RSSV_99}. We found that the quark constituent
mass $\Sigma\simeq 10$ MeV is significantly smaller than
the gap, $\Delta\simeq 50$ MeV. 

\section{The role of the strange quark mass}

  So far, we have only considered the case of three 
degenerate quark flavors. In the real world, the strange
quark is significantly heavier than the up and down 
quarks. The role of the strange quark mass in the high
density phase was studied in \cite{SW_99b,ABR_99}. The 
main effect is a purely kinematic phenomenon that is 
easily explained. The Fermi surface for the strange quarks
is shifted by $\delta p_F=\mu-(\mu^2-m_s^2)^{1/2}\simeq
m_s^2/(2\mu)$ with respect to the Fermi surface of the 
light quarks. The condensate involves pairing between
quarks of different flavors at opposite points on the 
Fermi surface. But if the Fermi surfaces are shifted, 
then the pairs do not have total momentum zero, and they
cannot mix with pairs at others points on the Fermi surface.
In a superfluid the Fermi surface is smeared out over a 
range $\Delta$. This means that pairing between strange 
and light quarks can take place as long
as the mismatch between the Fermi momenta is smaller
than the gap, 
\be
\label{m_s_crit}
 \Delta > \frac{m_s^2}{2\mu}.
\ee 
This conclusion is supported by a more detailed analysis
\cite{SW_99,ABR_99}. Since flavor symmetry is broken, we
allow the $\langle ud\rangle$ and $\langle us\rangle =
\langle ds\rangle$ components of the CFL condensate to be 
different. The $N_f=2$ phase corresponds to $\langle us
\rangle = \langle ds\rangle =0$. We find that there is 
a first order phase transition from the CFL to the $N_f=2$
phase. The critical strange quark mass is in rough agreement 
with the estimate (\ref{m_s_crit}). 

 This brings up the question whether QCD is in the 
CFL phase for realistic values of the quark masses and for 
physically relevant densities $\rho=\simeq (5-10)\rho_0$. This 
question is difficult to answer since it requires an accurate 
estimate of the gap and of the dynamically generated strange 
quark mass in the high density phase. We will see in the 
next section that for asymptotically large densities
the gap grows as a function of density. This means that
QCD will eventually enter the CFL phase for any value of 
the strange quark mass. 

\section{Superconductivity from perturbative gluon exchanges}

\begin{figure}[t]
\begin{minipage}[b]{80mm}
%\framebox[79mm]{\rule[-26mm]{0mm}{52mm}}
\epsfxsize=7cm
\epsffile{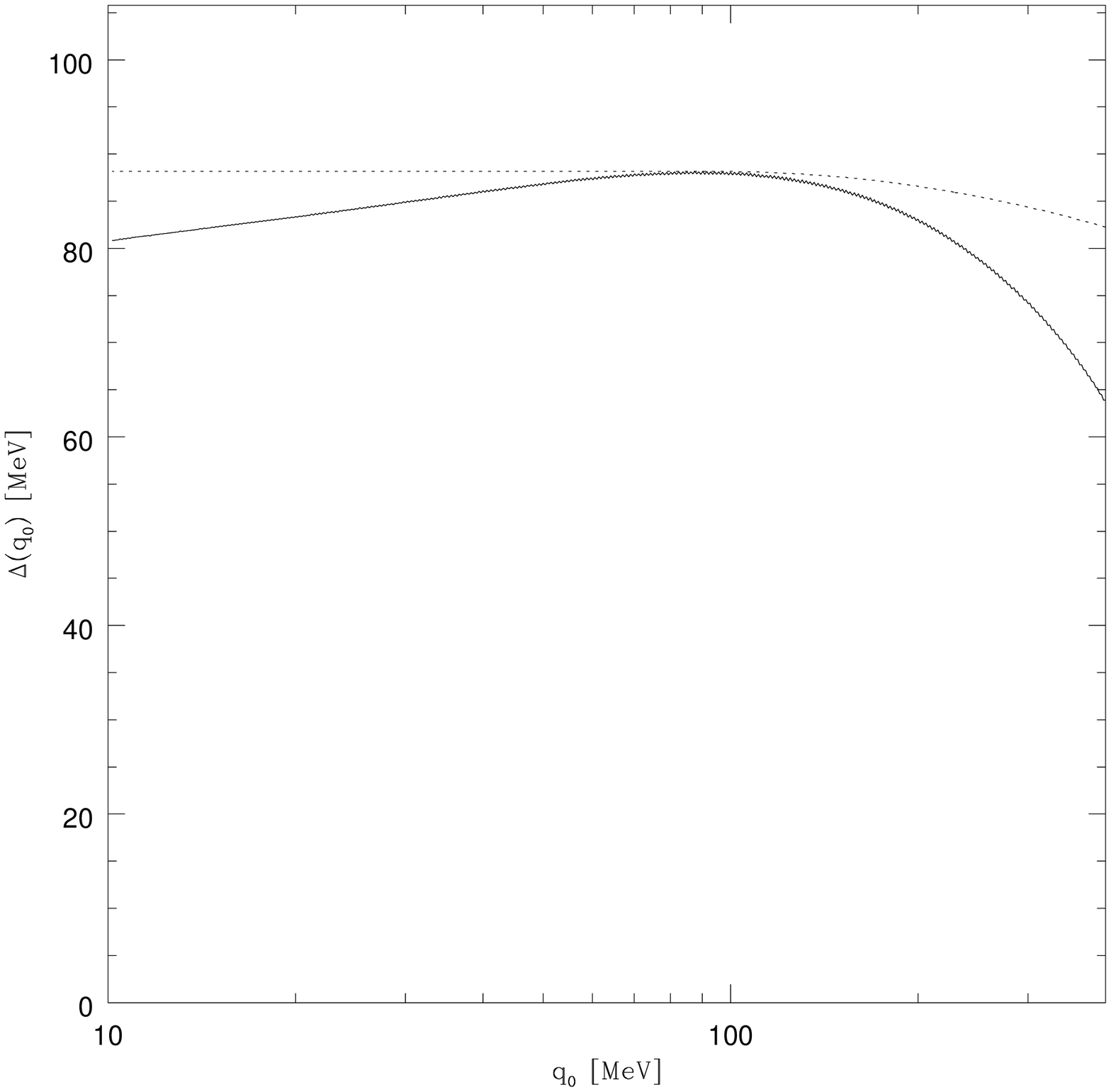}
\vspace*{-1cm}
\caption{\label{fig_eliash}
Solution of the Eliashberg equation with electric and magnetic
gluon exchanges for $\mu=400$ MeV.}
\end{minipage}
\hspace{\fill}
\begin{minipage}[b]{75mm}
%\framebox[74mm]{\rule[-26mm]{0mm}{52mm}}
\epsfxsize=7cm
\epsffile{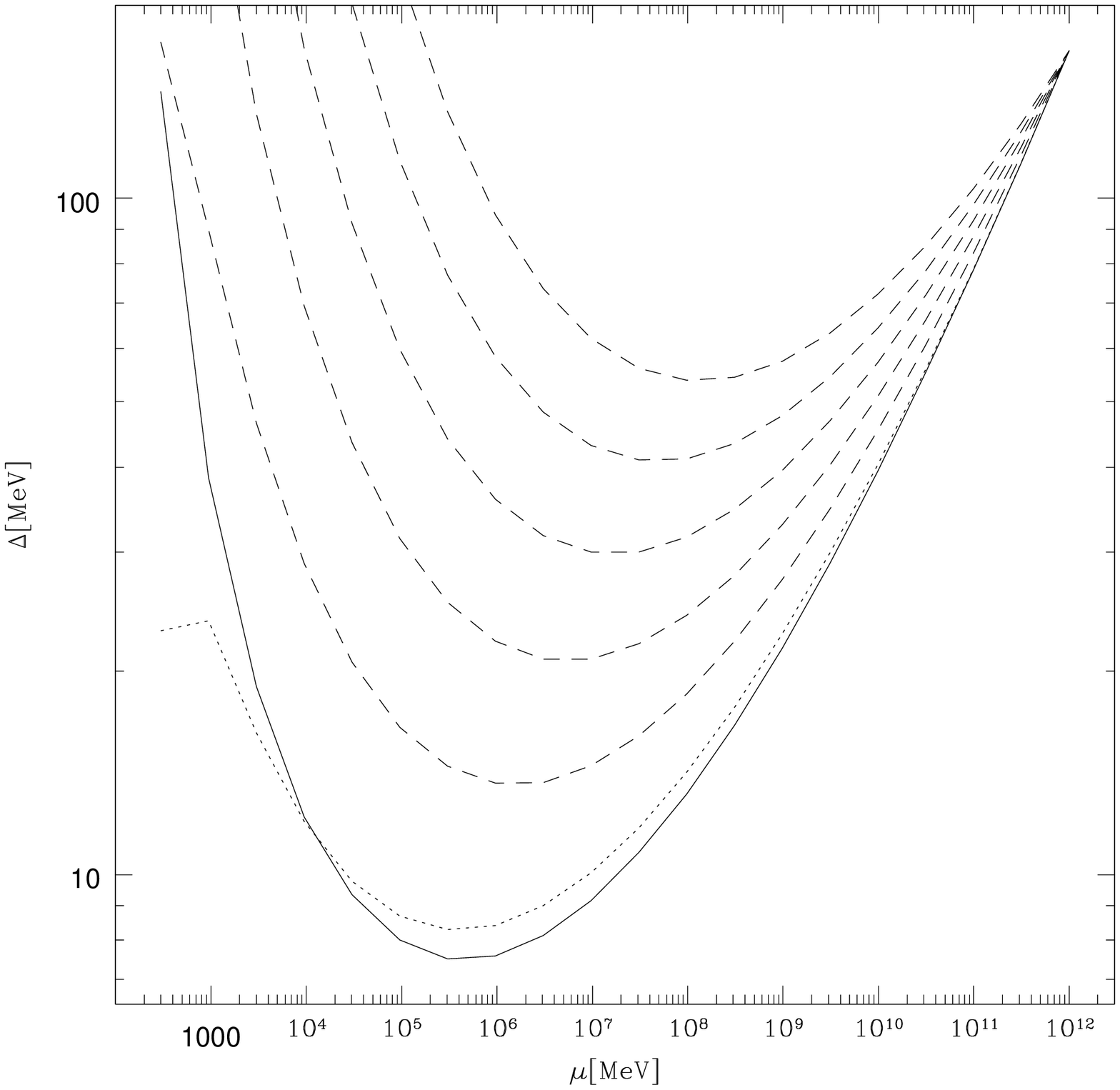}
\vspace*{-1cm}
\caption{\label{fig_scale}
Dependence of the gap on the chemical potential, compared
to a simple scaling form $g^{-k}\exp(-c/g),\;(k=1,\ldots 5)$. }
\end{minipage}
\end{figure}

 In our initial work the gap was calculated from an instanton
induced interaction. This interaction reproduces the strength of
chiral symmetry breaking at low density and leads to gaps on 
the order of 100 Mev at several times nuclear matter density. 
At very high density, instantons are suppressed and the gap
is dominated by perturbative gluon exchanges. This problem was
recently reexamined by Son \cite{Son_98}, who pointed out that 
unscreened magnetic gluon exchanges lead to a gap that scales 
as $\Delta\sim \exp(-const./g)$, rather than the naive expectation
$\Delta\sim\exp(-const./g^2)$. We have recently strengthened
this result by deriving an Eliashberg equation for the gap 
in the weak coupling limit \cite{SW_99b}. Including magnetic
gluon exchanges only, the gap equation reads 
\be
\label{eliash}
\Delta(p_0) = \frac{g^2}{18\pi^2} \int dq_0
 \log\left(1 + \frac{64\pi\mu}{N_fg^2|p_0-q_0|}\right) 
 \frac{\Delta(q_0)}{\sqrt{q_0^2+\Delta(q_0)^2}}.
\ee
This equation is independent of the gauge parameter in
a class of covariant gauges. The logarithm arises from 
almost colinear magnetic gluon exchanges. The colinear 
singularity is regulated by dynamic screening. The effects
of screening are taken into account by using the gluon
propagator in the hard dense loop approximation. Magnetic 
gluon exchanges generate a gap that scales as $\Delta
\simeq c\mu g^{-2}\exp(-3\pi^2/(\sqrt{2}g))$, where the 
coefficient in front of the exponent is on the order of 
$64\pi$ (for $N_f=2$).

 We have also included the effects of electric gluon exchanges. 
Electric gluons are less important than magnetic gluons because 
electric screening takes place at $q_E\sim g\mu$, while dynamic
screening sets in at $q_M\sim (g^2\mu^2\Delta)^{1/3}$. As a 
consequence, electric gluons do not modify the coefficient in the 
exponent, but they change the overall magnitude of the gap. 
A typical solution of the Eliashberg equation with both
electric and magnetic gluon exchanges included is shown 
in Fig. \ref{fig_eliash}. The scaling of the gap with 
the chemical potential is shown in Fig. \ref{fig_scale}. 
Here, we have used the one-loop running coupling constant
$g(\mu)$. The result is well described by
\be 
\Delta \simeq c\mu g^{-5}\exp(-3\pi^2/(\sqrt{2}g))
\ee
with $c\simeq 256\pi^4$. 
We find that for densities that are of physical interest, $\mu<
500$ MeV, the gap reaches $\Delta\simeq 100$ MeV. We should caution
that in this regime, $g\simeq (2-4)$ and the result may not be
reliable. Nevertheless, it is gratifying to see that the result
matches the number inferred from effective interactions at low
density.

\end{document}